\documentclass[12pt]{iopart}

\usepackage{iopams}
\usepackage{graphicx}

\begin{document}

\title{Classical analogy for the deflection of flux avalanches by a metallic layer}

\author{J Brisbois$^1$, B Vanderheyden$^2$, F Colauto$^3$, M Motta$^3$, W A Ortiz$^3$, J Fritzsche$^4$, N D Nguyen$^1$, B Hackens$^5$, O-A Adami$^1$ and A V Silhanek$^1$}

\address{$^1$ D\'epartement de Physique, Universit\'e de Li\`ege, B-4000 Sart Tilman, Belgium}
\address{$^2$ SUPRATECS and Department of Electrical Engineering and Computer Science, Universit\'e de Li\`ege, B-4000 Sart Tilman, Belgium}
\address{$^3$ Departamento de F\'{i}sica, Universidade Federal de S\~{a}o Carlos, 13565-905 S\~{a}o Carlos, SP, Brazil}
\address{$^4$ Department of Applied Physics, Chalmers University of Technology, S-412 96 G\"{o}teborg, Sweden}
\address{$^5$ NAPS/IMCN, Universit\'{e} catholique de Louvain, B-1348 Louvain-la-Neuve, Belgium}
\ead{jbrisbois@ulg.ac.be}

\date{\today}

\begin{abstract}
Sudden avalanches of magnetic flux bursting into a superconducting sample undergo deflections of their trajectories when encountering a conductive layer deposited on top of the superconductor. Remarkably, in some cases flux is totally excluded from the area covered by the conductive layer. We present a simple classical model that accounts for this behaviour and considers a magnetic monopole approaching a semi-infinite conductive plane. This model suggests that magnetic braking is an important mechanism responsible for avalanche deflection.

\end{abstract}


\maketitle

\section{Introduction}

Faraday's concept of lines of flux emanating from magnets provides a pedagogical way to visualize magnetic or other vectorial fields. Even though these field lines represent a mere mathematical construction, in type-II superconductors where a continuous field breaks up in small tubes of quantized units of flux, they are close to acquire physical significance. In principle, these superconducting flux lines (or vortices) penetrate through the sample's borders and rush to the center of the superconductor as soon as a magnetic field is applied. However, inevitable and ubiquitous sample imperfections impede the motion of vortices and give rise to a gradient distribution of magnetic field  given by $\nabla \times \bf{B}= \mu_0 \bf{J_\mathrm{c}}$, where $\mu_0$ is the vacuum permeability and $\bf{J_\mathrm{c}}$ the critical current density. This so-called critical state is metastable and therefore prone to relax to the equilibrium state corresponding to a more homogeneous field distribution. The relaxation process can be achieved via thermal activation of flux bundles \cite{anderson62,anderson64} over the pinning potential landscape (flux-creep) if a fast thermal diffusion allows an efficient removal of the heat produced by flux hopping, thus keeping the superconductor under isothermal conditions. A different scenario arises when thermal diffusion to the surrounding is slow. Under these circumstances, local heating leads to a reduction of the critical current, which in turn favours further vortex displacement and heat production \cite{mints}. This positive feedback loop eventually triggers a jet of flux lines bridging the border and the center of the sample in a very short time. Clearly, these thermomagnetic instabilities and the over-heated trail they leave behind can have very detrimental or even catastrophic consequences in technological superconducting applications, as seen for instance in the spectacular quenching of superconducting magnets \cite{parks}. 

It was already noted in early days that copper coating of superconducting solenoids provided a simple remedy to increase the thermal diffusion and consequently, to decrease the thermomagnetic instabilities \cite{parks}. Later on, similar suppression of dendritic flux avalanches have been observed in superconducting thin films with a metallic capping layer and naturally attributed to their improved thermal-sink effect \cite{baziljevich02,Choi-APL05,Choi-SUST09}. It was only recently that an alternative mechanism, other than thermal, has been invoked to explain the suppression of flux jumps. First, Albrecht {\it et al.} \cite{albrecht05} noticed that avalanches propagating into an Au-covered region change the propagation direction depending on the incident angle. This observation led the authors to conclude that large electric fields induced in the Au are responsible for these avalanche deflections and that avalanches propagate at slower velocity under the Au layer. In 2010, Colauto {\it et al.} \cite{colauto10} provided unambiguous confirmation of Albrecht's interpretation when reporting on the suppression of avalanches even if the metallic layer is located far apart from the superconductor. These findings pointed out the relevance of the magnetic braking of flux motion caused by induced eddy currents in the metallic layer and questioned the hypothesis of phonon escaping through the conductive layer.       

Interestingly, the problem of increased damping of superconducting vortices when moving under a conductive layer had been already experimentally and theoretically addressed by Rojo and co-workers. Indeed, Danckwerts {\it et al.} \cite{danckwerts00} observed an additional damping of vortex motion in a superconductor/semiconductor hybrid system caused by the eddy currents in the \textsc{2D} electron gas. By changing the number of carriers with a voltage gate on the \textsc{2D} electron gas, the vortex damping could be controlled. A theoretical analysis of this phenomenon was performed by Baker and Rojo \cite{baker01} for a single vortex and for a chain of vortices. A more macroscopic study of the influence of inductive braking on the morphology of avalanches has been carried out in \cite{vestgaarden13,vestgaarden14}.

In this work we present experimental evidence, via magneto-optical imaging, that a conductive layer (Cu) can repel flux avalanches triggered in a underlying superconducting film (Nb). By placing the conductive layer away from the borders of the superconducting film, we ensure no influence of the Cu on the early development of the thermomagnetic instabilities and guarantee that the flux avalanche is running at high speed when entering into the surface covered by the Cu \cite{fullspeed1,fullspeed2}. We then address the question of whether a single vortex driven by a constant force would undergo a deflection of its trajectory when encountering a metallic layer, assuming isothermal conditions. Using a classical analogy where the vortex is substituted by a magnetic monopole \cite{carneiro}, we demonstrate that (i) Baker and Rojo's calculations of vortex damping need to be corrected at high vortex velocities, and (ii) trajectory deflection and even total repulsion of the monopole should take place. More precisely, we show that the conductive layer gives rise to a non-monotonous damping force similar to that caused by vortex contraction or vortex expansion at high vortex velocities \cite{LO,kunchur,gurevich}. Since the indicator films typically employed in magneto-optical imaging setups include an aluminium mirror of about $100$ nm thick which, in turn, is positioned in close proximity to the surface of the superconductor, one should always bear in mind that, under certain circumstances, the 
general assumption that magneto-optical imaging is a non-invasive technique might not be entirely valid.

\section{Experimental results}

\begin{figure*}[ht!]
\includegraphics[width=\linewidth]{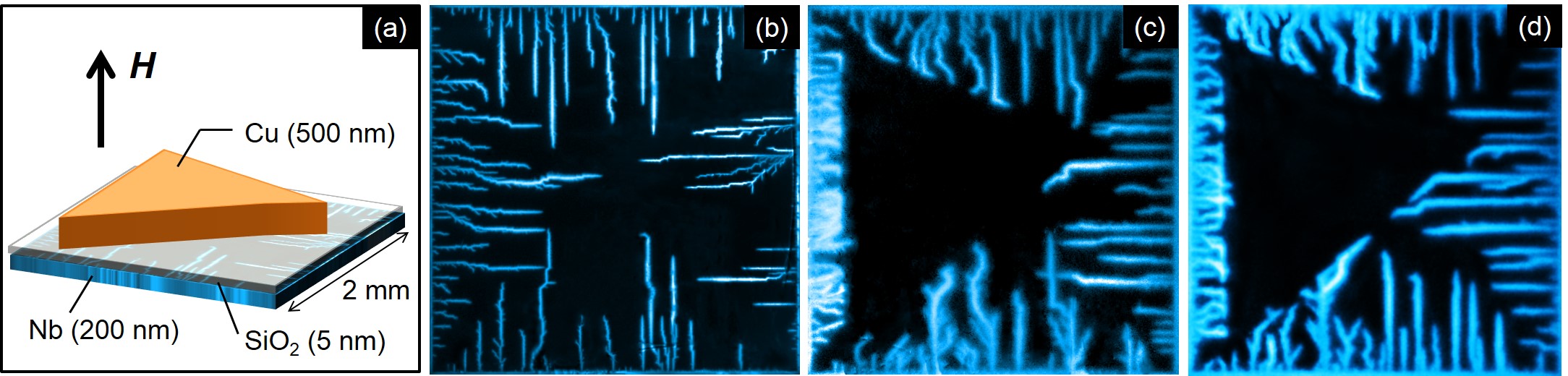}
\caption[]{(a) Sketch of the sample layout. Panels (b) and (c) show magneto-optical images taken at $T=2.5$ K after zero-field cooling in  $H=10$ Oe for the Nb sample before (b) and after (c) covering it with the Cu triangle.  The bright areas correspond to the highest magnitudes of the magnetic field, while the dark areas represent the lowest fields. Panel (d) shows a MOI picture of the Nb sample with the Cu triangle after field-cooling in $H=10$ Oe down to $T=2.5$ K and subsequently turning off the magnetic field. In (d) the color code has been reversed such that antivortex avalanches appear as brighter.} 
\label{fig:nbcu_avalanches}
\end{figure*}

The sample consists of a $50$ nm thick Nb film of lithographically-defined square shape with $2$ mm side length. In order to increase the $H-T$ region where thermomagnetic instabilities occur \cite{menghini05}, we have patterned the Nb film with a periodic square array of antidots of 4 $\mu$m pitch and antidot size of 1.5 $\mu$m. The superconducting critical temperature was $T_\mathrm{c}=8.3$ K, estimated coherent length $\xi(0) \sim 12$ nm, and penetration depth $\lambda(0) \sim 92$ nm \cite{motta14}. 
The magnetic flux distribution of the as-fabricated system was measured by magneto-optical imaging (\textsc{MOI}), based on the field-dependent rotation of the light polarization in an indicator film placed on top of the superconducting specimen \cite{moimaging}. The indicators used in the present work are Bi-substituted yttrium iron garnet films (Bi:YIG) with in-plane magnetization. In a subsequent process step, a $500$ nm thick copper layer of triangular shape defined by electron-beam lithography was evaporated on top of the Nb sample. A $5$ nm thick SiO$_2$ layer, prepared using chemical vapour deposition, separates the Nb and Cu films to avoid proximity effects. The Cu triangle was purposely placed far from the sample's borders in order to avert any risk of thermal shunt at the nucleation point of the avalanches. Figure \ref{fig:nbcu_avalanches}(a) shows a schematic representation of the final sample layout.


The most important result is summarized in figure \ref{fig:nbcu_avalanches} and consists of a clear exclusion of flux avalanches by the Cu layer, as evidenced by comparing the flux entrance for the sample without Cu [figure \ref{fig:nbcu_avalanches}(b)] and with Cu [figure \ref{fig:nbcu_avalanches}(c)] in zero-field cooling conditions at $T=2.5$ K and $H=10$ Oe. Similar effect is observed if the sample is first field-cooled in $H=10$ Oe down to $T=2.5$ K and imaged at remanence after turning off the field  [figure \ref{fig:nbcu_avalanches}(d)]. In this latter case, the sample is initially full of vortices and the avalanches correspond to anti-vortices penetrating from the border of the sample. In all cases, the images unambiguously show that the avalanches avoid entering the area covered by the Cu layer and are deflected along its perimeter. At higher fields, the flux dendrites penetrate into the triangular area following the main directions of the underlying pinning lattice, similarly to the behaviour recently shown in \cite{motta14}. Irrespective of the applied field intensity, in the avalanche regime the mean field value under the Cu-layer remains smaller than in the rest of the Nb film. The images corresponding to the bilayer Nb/Cu system [panels (c) and (d)] show a somewhat lower resolution than the bare Nb film due to the fact that the Cu spacer places the indicator further away from the Nb surface. Images taken in the smooth (critical state) flux penetration regime show no difference between the sample with or without the Cu triangle. This observation clearly indicates that the velocity of flux propagation plays an important role on the deflection of avalanches, and the here reported phenomenon could be thought of as a \textsc{2D} skin depth effect.

\section{Classical model}

\begin{figure}[ht!]
\centering
\includegraphics[width=\linewidth]{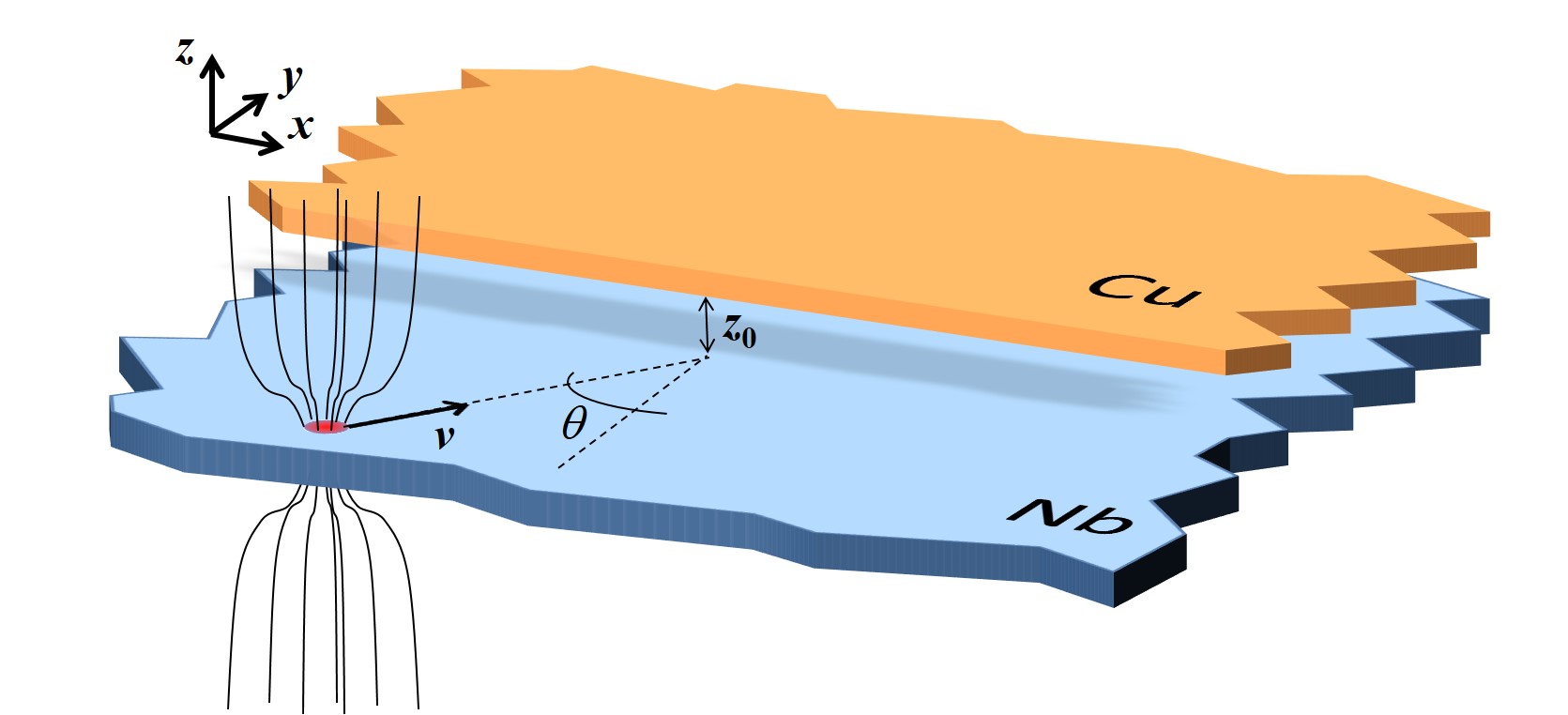}
\caption[]{Scheme of the classical problem discussed in the text. A single magnetic monopole (vortex) propagating in an overdamped medium (Nb) is pushed towards a conducting layer (Cu) with a constant driving force $F_0$ forming an angle $\theta$ with the normal to the border of the conducting layer. The metallic sheet is separated by a distance $z_0$ from the plane containing the magnetic charge.} 
\label{fig:initial_sit}
\end{figure}

A macroscopic description of the flux avalanches depicted in figure \ref{fig:nbcu_avalanches} requires the coupling of heat transport equations and Maxwell equations with the constitutive relation corresponding to the superconducting state and the inductive link to the conductive layer. This approach seems justified since avalanches involve a large number of vortices and therefore knowledge of the behaviour at the single vortex level may not be required. Moreover, numerical modelling suggests that the local temperature in the avalanche trail can rise above the superconducting critical temperature and therefore an avalanche can be better pictured as a propagating normal/superconductor interface rather than a moving vortex bundle \cite{motta14}.  All in all, it is undoubtful that an avalanche can be essentially described as a propagating magnetic flux front. This brings up the question of how this travelling magnetic flux front will interact with a conductive layer. In particular, we wonder whether a single individual vortex would undergo any detour of its initial trajectory when penetrating the region covered by the conductive layer. In this section we answer this question by analysing a classical problem of a magnetic monopole, emulating an individual vortex travelling in an overdamped medium, and its interaction with a conductive plate, as illustrated in figure \ref{fig:initial_sit}.

\subsection{Infinite conducting plane}

Let us start reviewing the general problem of the forces acting on a magnetic charge (monopole) when it moves in the vicinity of a conducting plane of infinite spatial extension. We will discuss the more complicated situation involving the border of the conductor in the following section.

\begin{figure}[ht!]
\centering
\includegraphics[width=\linewidth]{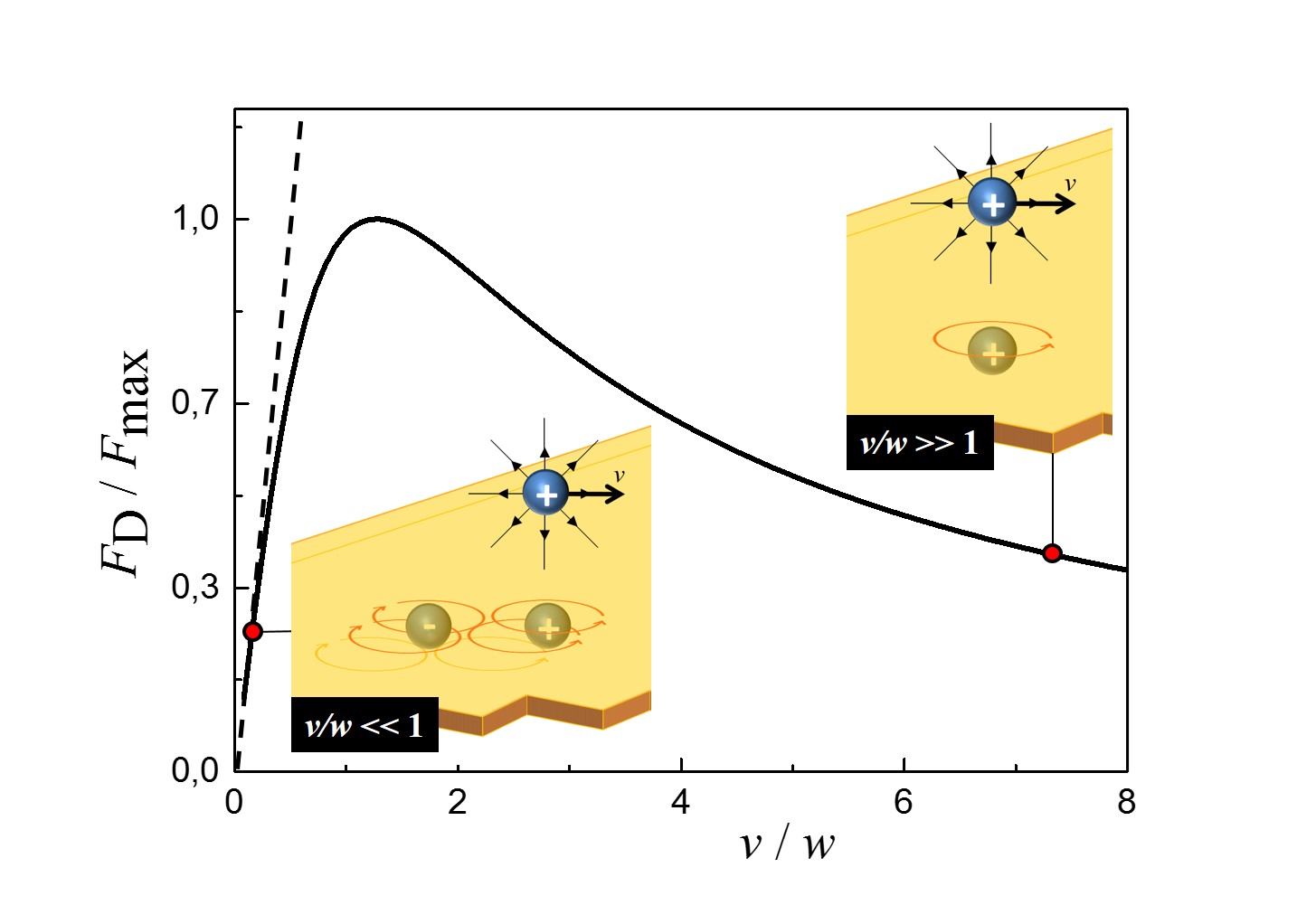}
\caption[]{Comparison of the models developed by Baker-Rojo  \cite{baker01} (dotted line) and that of Reitz \cite{reitz70} (solid line) for the drag force acting on a moving monopole in presence of a conducting layer. The insets illustrate the current images induced in the two extreme situations by the moving monopole with positive magnetic charge. A similar picture is obtained for a moving dipole \cite{reitz70,liu}.} 
\label{fig:drag_force}
\end{figure}

As early as in 1872, J.C. Maxwell already discussed the induction of electric currents in an infinite plane sheet of uniform electrical conductivity $\sigma$ by a moving magnet \cite{maxwell1872}. In few words, due to Faraday's induction law, when a monopole of positive magnetic charge travels at a distance $z_0$ on top of a conductive plane, it induces counterclockwise swirls of eddy currents ahead of the moving magnet (when seen from above) and a clockwise loop of eddy currents lagging behind the magnet. The magnetic field generated by these eddy currents is equivalent to that produced by a negative image of the monopole on its trailing edge and a positive image on the leading edge, both images situated at a distance $-z_0$ from the conducting plane. Due to the finite conductivity of the metallic sheet, these induced images (or eddy currents) gradually disappear which is equivalent to say that the images propagate downward at a speed $w \propto \sigma^{-1}$. Naturally, if the conductivity is infinite, i.e. a superconductor, these currents will not fade out. The above considerations for a monopole can be used as building blocks for an arbitrary multipolar distribution. In particular, in the case of a dipole with magnetic moment in the $+z$ direction, the eddy current patterns are very similar to those induced by a moving monopole. When considering the forces acting on the magnet we should add the interaction of the infinite trail of images that the magnet is leaving behind its path (see lower inset in figure \ref{fig:drag_force}). A more modern and pedagogical description of this problem can be found in \cite{reitz70,saslow,saslow91,rossing,liu,lee,thess}.

At low magnet velocities $v \ll w$, only the new induced positive and negative images matter, since the others have receded long before. Since the leading image is positive and the trailing image is negative, both hold back the monopole, leading to a damping force known as magnetic braking. In this limit, the drag force $F_\mathrm{D}$ is proportional to $v$ as for a viscous medium. Interestingly, a different scenario appears at high velocities such that $v \gg w$. Now, during the time the magnet moves forward the images have not receded significantly and therefore they cancel out in the limit of infinite velocity or conductivity, leaving only the positive image. Since magnet and image are facing the same magnetic poles, they repel each other giving rise to a levitation force on the magnet whereas the drag force tends to diminish (see upper inset in figure \ref{fig:drag_force}). As was already pointed out in \cite{reitz70,liu}, similar effects should be observed irrespective of whether we deal with a magnetic monopole or a dipole perpendicular to the conducting plane.


The general expression for the drag force is \cite{reitz70}

\begin{equation}\label{eq:fd_reitz}
F_\mathrm{D} = \frac{\pi C}{z_0^2} \frac{w}{v} \left( 1- \frac{w}{\sqrt{v^2+w^2}} \right),
\end{equation}
where $C$ is a constant equal to $C_\mathrm{m} = \mu_0 q^2/16 \pi^2$ for a monopole, $q$ is the magnetic charge of the monopole and $z_0$ is the distance between the monopole and the conductor. In the case of a dipole, $C$ is given by $C_\mathrm{d} = 3 \mu_0 m_\mu^2/32 \pi^2 z_0^2$, where $m_\mu$ is the magnetic moment of the dipole.

The lift force is given by $F_\mathrm{L} = (v/w) F_\mathrm{D}$ and increases monotonously with $v$. This force is of no particular  importance in our treatment of the deflections of vortex trajectories. Figure \ref{fig:drag_force} shows the dependence of $F_\mathrm{D}$ with velocity and an illustration of the magnetic images in the limiting cases.

It is instructive to compare the analysis presented above to the calculation by Baker and Rojo \cite{baker01} of the viscosity in the case of a superconducting vortex moving close to a two-dimensional electron gas (\textsc{2DEG}) with conductivity $\sigma_\mathrm{2DEG}$. These authors considered the electric field $\bf{E}$ induced in the \textsc{2DEG} by the varying magnetic field of the moving vortex and deduced an expression for the power $\sigma_\mathrm{2DEG} {\bf E}^2$ dissipated by the eddy currents as a function of  the vector potential ${\bf A}(\bf{r})$ associated with the vortex field. This leads to a drag force ${\bf F}_\mathrm{D}=-\eta_\mathrm{2DEG} \bf{v}$, with the viscosity $\eta_\mathrm{2DEG}$ associated with the \textsc{2DEG} given by
\begin{equation}
\eta_\mathrm{2DEG}= \alpha(d,z_0,\kappa) \sigma_\mathrm{2DEG} \frac{\Phi_0^2}{4\pi \lambda^2},
\end{equation}
where $\Phi_0$ is the fundamental quantum of flux and $\alpha$ is a constant depending on the thickness of the superconductor $d$, the distance $z_0$ between the superconductor and the \textsc{2DEG} and the Ginzburg-Landau parameter $\kappa=\lambda/\xi$. The additional viscosity  $\eta_\mathrm{2DEG}$ should be added to the Bardeen-Stephen \cite{bardeen} viscosity $\eta_\mathrm{SC}$ produced by the vortex motion in the superconductor. Baker and Rojo showed that for a \textsc{2DEG}, normally $\eta_\mathrm{SC} \gg \eta_\mathrm{2DEG}$.

It is worth noting that Baker and Rojo predict a constant damping coefficient, whereas Maxwell's analysis leads to a non-monotonous $\eta(v)$  \cite{nonnewtonian}. The reason for this apparent discrepancy is that the Baker-Rojo approach is only justified at low velocities, when the magnetic flux lines have the time to fully penetrate into the metallic layer \cite{rossing} whereas, at high velocities, the influence of the magnetic field generated by the eddy currents on the moving vortex cannot be neglected anymore. The response of the system can also be described in terms of an electric circuit \cite{saslow}. At low velocities, the conductor perceives low-frequency variations of the source magnetic field. In this case, the induced currents are weak and their contribution to the total magnetic field is negligible, so the system is mostly resistive. At higher frequencies however, the magnetic field generated by the eddy currents counteracts the source fields and can no longer be neglected. The flux lines are expelled from the conductor and the system response is subsequently dominated by self-inductance effects. 

The fact that at a certain critical velocity  $v_c \simeq 1.27 w$ the drag force becomes smaller as $v$ increases, resembles the non-linear damping of individual vortices moving faster than an instability velocity $v^*$ as described by Larkin and Ovchinnikov \cite{LO}. As we pointed out above, for a vortex moving in a superconductor capped with a metallic layer, both damping coefficients should be added and consequently two instability points will appear. The critical velocity $v_\mathrm{c}$ is determined by the relaxation time of the electrons in Cu, whereas  $v^*$ is given by the relaxation time of quasiparticles in Nb. In our particular case of a $500$ nm thick Cu layer, we find that $v_\mathrm{c} \sim 100$ m/s, whereas for Nb $v^* \simeq 100-1000$ m/s \cite{grimaldi}. These values should be compared with the velocity of avalanches in thin superconducting layers \cite{fullspeed2} which can attain $10$ km/s. To our knowledge there is no report so far on the influence of a metallic layer on the Larkin-Ovchinnikov instability. Knowing that the Larkin-Ovchinnikov instabilities are the precursors for the development of phase-slip lines \cite{vodolazov,silhanek,vandevondel}, it would be also interesting to explore the delay of formation of phase slip lines and hot spots in metal-coated superconductors.

\subsection{Influence of the border}

\begin{figure}[ht!]
\includegraphics[width=\linewidth]{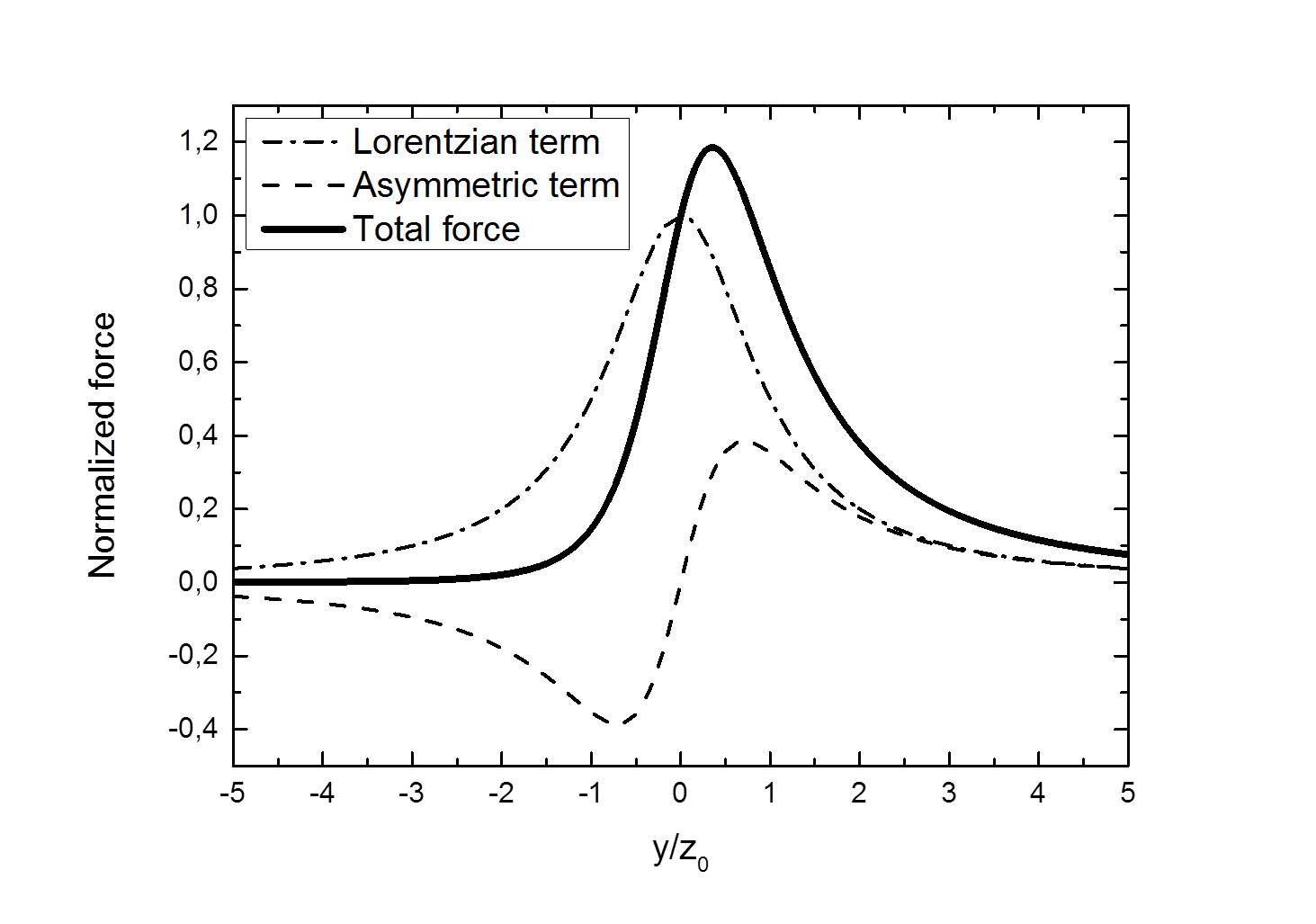}
\caption[]{Contribution of the two terms composing the lateral force tending to push a magnet away from a conducting plane, for $C/z_0^2=1$. In our simple model (subsection \ref{subsec:model}), we neglect the contribution of the asymmetric term and work on the main Lorentzian term.} 
\label{fig:flat}
\end{figure}

We considered so far the case of an infinitely extended plane.
Incorporating the effects of borders is highly non-trivial, as discussed by Davis and co-workers, who considered a monopole moving over a conducting plane of semi-infinite extension \cite{davis71,borcherts}. The receding image technique cannot be applied in this situation and no analytic solution is known. However, by using conformal mapping techniques, Davis and co-workers managed to obtain an analytic solution in the limit of an infinite electrical conductivity. Since the diffusion velocity $w$ is $\sim \sigma^{-1}$, the limit $\sigma \to \infty$ can also be seen as that of an infinite monopole velocity, with $v \gg w \to 0$ \cite{davis71,borcherts}. In such case, the monopole experiences, additionally to the drag force, a lateral force $F_\mathrm{lat}$ which pushes it away from the region covered by the perfectly conducting plane. The effect can be understood as a consequence of the asymmetric eddy current distribution which is compressed by the border of the conductor \cite{davis71}. An analytical expression for $F_\mathrm{lat}$ was calculated for a magnetic monopole close to the border of a perfect conductor, which is equivalent to the high velocity limit of a monopole moving parallel to the border \cite{davis71}. If the monopole is at a vertical distance $z_0$ from a semi-infinite conducting plane $y>0$, the force can be written
\begin{equation}
{\bf F}_\mathrm{lat}= - C_\mathrm{m} \left( \frac{1}{y^2+z_0^2} + \frac{y}{(y^2+z_0^2)^{3/2}} \right) \hat{y}.
\end{equation}
The first term corresponds to a symmetric Lorentzian peak centred at $y=0$, while the second term is asymmetric. The general shape of ${\bf F}_\mathrm{lat}(y)$, as represented on figure \ref{fig:flat}, is thus an asymmetric peak with a maximum located inside the conducting plane.  The maximum force is $F_\mathrm{max}=-32C/27z_0^2$, the position of the maximum is $y_\mathrm{max}=z_0/2\sqrt{2}$ and the half-height width is $W_{1/2}=2.18 z_0$.

\subsection{Vortex trajectories} 
\label{subsec:model}

Based on the aforementioned results we will now develop a model describing the trajectory of a vortex when penetrating the region covered by the metallic layer. To tackle this problem, we approximate the magnetic field of a vortex by a magnetic monopole \cite{carneiro}. We consider the situation depicted on figure \ref{fig:initial_sit} where a magnet is launched from $y<0$ towards a semi-infinite plane with perfect conductivity located at $y>0$. Assuming perfect conductivity ensures a maximum effect of the metallic layer on the magnet. Let us consider the case where the magnet is pushed by a constant force ${\bf F}_0= \eta {\bf v}_0$, where $\eta$ is the viscosity of the medium. The initial velocity ${\bf v}_0$ has an angle $\theta$ with respect to the normal to the $y=0$ interface. 

The magnet is thus moving in a highly viscous medium where inertia plays no role at all, meaning that the response is determined by the forces exerted at the moment and by nothing in the past. This approximation is fully justified in the case of a superconducting vortex where inertial terms are known to be very small \cite{nomass}. Moreover, we will neglect the drag force ${\bf F}_\mathrm{D}$ (due to the eddy currents induced in the conductor) opposed to the velocity ${\bf v}$, as it will only slightly affect the trajectories and merely change the effective modulus of ${\bf F}_0$. This assumption is particularly valid in the limit $\eta \gg \eta_\mathrm{2DEG}$ described by Baker and Rojo.

As we anticipated above, due to the presence of the conductor, the magnet will experience a lateral force ${\bf F}_\mathrm{lat}(y)$ perpendicular to the interface. In order to obtain simple analytical expressions for the vortex trajectories, we will retain only the dominant symmetric term of  ${\bf F}_\mathrm{lat}(y)$ (see  figure \ref{fig:flat}). Using the notations of figure \ref{fig:initial_sit}, and writing $v_{0,x} = v_0 \cos \theta$ and $v_{0,y} = v_0 \sin \theta$, the equations of motion can be expressed as follows:
\begin{equation}
F_x = \eta v_{0,x} = \eta \frac{\rmd x}{\rmd t},
\end{equation}
\begin{equation}
F_y = \eta v_{0,y} + F_\mathrm{lat}(y) = \eta \frac{\rmd y}{\rmd t}.
\end{equation}
By combining these two equations, we obtain the equation for the magnet trajectory:

\begin{equation}\label{eq:trajmod}
x(y) = x_0 + \int_{y_0}^y \rmd y' \frac{y'^2+z_0^2}{a y'^2 + a z_0^2 - b},
\end{equation}
where we define the parameters $a \equiv v_{0,y} / v_{0,x}$ and $b \equiv C / \eta v_{0,x}$.

From this equation, we can distinguish three cases, depending on the values of $a$, $b$ and $z_0$.

\begin{figure}[ht!]
\includegraphics[width=\linewidth]{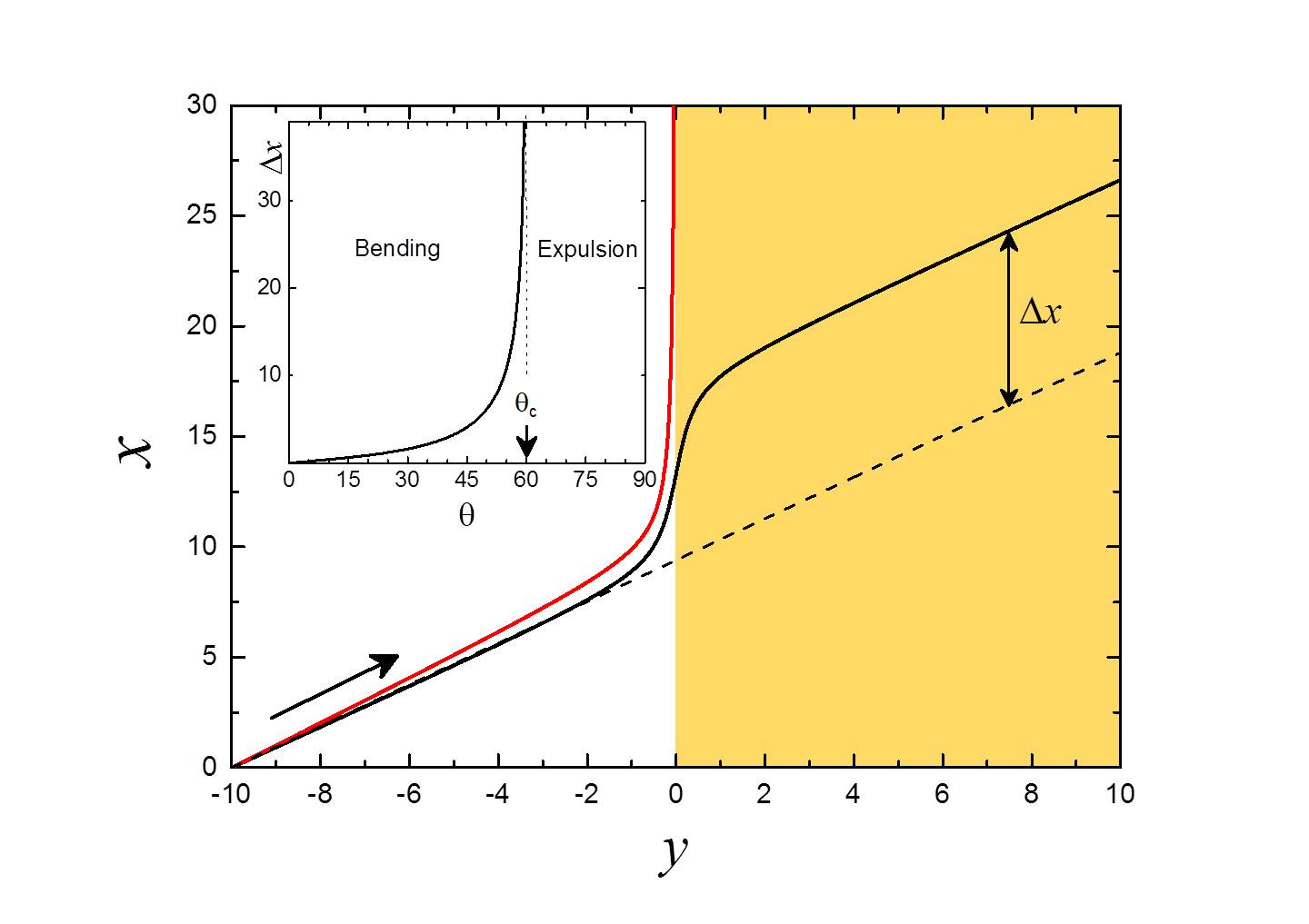}
\caption[]{Trajectory $x(y)$ of a magnet launched from $x_0=0$, $y_0=-10$ towards a conducting plane for $b=1$, $z_0=1$ and (red) $a=1$; (black) $a=1.1$. The deflection is bigger as $a$ decreases towards the critical value $a=b/z_0^2 = 1$. For smaller values of $a$, the magnet is completely repelled by the conductor. Inset: $\Delta x$ as a function of the initial angle $\theta$ for $C=0.5$, $F=1$ and $z_0=1$. The critical angle $\theta_\mathrm{c}$ above which the magnet does not penetrate in the region covered by the conducting layer is indicated. Bending (exclusion) of the vortex trajectory corresponds to the black (red) line in the main panel. The $x$ and $y$ coordinates are normalized by $z_0$.} 
\label{fig:traj1}
\end{figure}

If $z_0^2 > b/a$, i.e. when the magnet is launched close to the normal direction, we can integrate by substitution with respect to $u = y/\sqrt{z_0^2 - b/a}$. By doing this, we obtain the trajectory:
\begin{equation}\label{eq:traj1}
x(y) = x_0 + \left. \frac{1}{a} y \right|_{y_0}^y + \frac{b}{a^2 \sqrt{z_0^2 - \frac{b}{a}}} \left. \arctan \left( \frac{y}{\sqrt{z_0^2 - \frac{b}{a}}} \right) \right|_{y_0}^y.
\end{equation}
When crossing the interface, the magnet is deflected in the direction of $\vec{v}_{0,x}$, as shown by the black line on figure \ref{fig:traj1}. From this equation, we can extract the amplitude of the deflection $\Delta x$ given by the difference between the third term of (\ref{eq:traj1}) at $y \rightarrow +\infty$ and $y \rightarrow -\infty$:
\begin{equation}
\Delta x = \frac{\pi b}{a^2 \sqrt{z_0^2-\frac{b}{a}}}.
\end{equation}

For given $z_0$ and $b$, the shift from the original direction, $\Delta x$, diverges when the angle of incidence of the magnet surpasses the critical angle

\begin{equation}
\theta_\mathrm{c} = \arccos \frac{C}{F z_0^2}.
\end{equation}
The behaviour of $\Delta x (\theta)$ is shown on figure \ref{fig:traj1}.

For $z_0^2=b/a$, the trajectory is given by
\begin{equation}
x(y) = x_0 + \left. \frac{y}{a} \right|_{y_0}^y - \left. \frac{b}{a^2} \frac{1}{y} \right|_{y_0}^y.
\end{equation}
In this case the magnet never crosses the interface as the lateral force compensates exactly the driving force and therefore the final trajectory approaches asymptotically the border of the conducting plane following the red line in figure \ref{fig:traj1}.

Finally, when $z_0^2 < b/a$ we obtain similar trajectories as that for $z_0^2=b/a$, with the magnet running parallel to the interface but at finite distance from it. This trajectory resembles the one followed by vortex avalanches shown in figure \ref{fig:nbcu_avalanches} thus suggesting that Faraday's induction law is the responsible for the observed exclusion of flux avalanches by the Cu layer.

A more precise calculation evaluating numerically the full expression of $F_\mathrm{lat}$ shows essentially the same results described above. Note that strictly speaking, the analysis based on Maxwell's receding images theory for an infinite plane is only valid when the skin depth $\delta$ is much bigger than the thickness of the metallic layer, $d \ll \delta$. However, a more complete analysis shows that the results are qualitatively the same in the high-frequency regime when $d \gg \delta$ \cite{reitz72}. In addition, in our analysis we have also neglected the fact that the magnetic images induced in the conducting layer by the moving vortex will, in turn, generate images on the superconducting layer. We would like to emphasize the fact that our classical model does not intend to \textit{quantitatively} account for the deflections of avalanches, but to point out that, even without invoking thermal effects, Faraday's law alone should give rise to deflections of travelling magnetic flux. In our approximation, we assumed a constant force acting on the vortex. This condition can be experimentally realized when the system is driven into the free flux flow regime by applying an external current, but will certainly not be accurate enough to account for the case of vortices driven by thermomagnetic instabilities. Another important consideration is the fact that our analysis deals with a single vortex and therefore neglects collective effects. Indeed, as soon as vortices enter the region covered by the metal, they slow down and tend to accumulate at the interface, thus developing a vortex dam which impedes the motion of new incoming vortices. This effect may lead to a substantial reinforcement of vortex deflection or to a reorientation of the flux front trajectory at the interface with the metal, very much like refraction of a light beam when traversing two media with different refractive indices and somewhat similar to the experimental results reported by Albrecht {\it et al.} \cite{albrecht05}.  

\section{Conclusion}

Motivated by the experimental observation of the exclusion of magnetic flux avalanches in a Nb sample  partially covered by a conducting capping layer, we have investigated the simplified case of the  interaction of a magnetic charge (monopole and dipole) with a semi-infinite conducting plane. We have found that early theoretical descriptions for the vortex damping enhancement due to the metallic sheet needed a correction at large vortex velocities where a decrease of the damping coefficient is expected. We also demonstrate that vortex trajectories are strongly modified when penetrating into the area covered by the metallic sheet and may even be fully diverted from that area thus providing a qualitative explanation for the bending of the trajectories of flux avalanches. Considering that typical magneto-optical experiments with yttrium iron garnet films need an aluminium mirror of about $100$ nm thick in close proximity to the surface of the superconductor, we question the general assumption that these measurements do not influence the experimental results. Our findings may be extended to study the damping of Larkin-Ovchinnikov vortex instabilities and phase-slip lines in current driven systems.

\ack This work was partially supported by the Fonds de la Recherche Scientifique - \textsc{FNRS}, the \textsc{ARC} grant 13/18-08 for Concerted Research Actions, financed by the Wallonia-Brussels Federation, the Brazilian National Council for Scientific and Technological Development (\textsc{CNP}q) and the S\~ao Paulo Research Foundation (\textsc{FAPESP}), Grant No. 2007/08072-0, and the program for scientific cooperation \textsc{F.R.S.-FNRS-CNP}q. J.B. acknowledges support from \textsc{F.R.S.-FNRS} (Research Fellowship). The work of A.V.S. is partially supported by "Mandat d'Impulsion Scientifique" of the F.R.S.-FNRS. B.H. acknowledges support from \textsc{F.R.S.-FNRS} (Research Associate). We would also like to thank X. Baumans, R. Delamare, D. do Carmo, V. Gladilin, T. H. Johansen, J. Lombardo and C.C. Souza Silva for useful discussions at different stages of the project development.

\end{document}